\documentclass[final]{elsart}
\usepackage{epsfig}
\newcommand{\m}{\phantom{-}}	
\newcommand{\z}{\phantom{0}}	
\newcommand{\MSbar}{$\overline{\mbox{MS}}$} 
\newcommand{\ie}{{\em i.e.}}

\newcommand{\etal}{{\em et al.}}
\newcommand{\eqref}[1]{Eq.~(\ref{#1})}   
\newcommand{\citeref}[1]{Ref.~\cite{#1}} 
\newcommand{\figref}[1]{Fig.~\ref{#1}}   
\newcommand{\tabref}[1]{Table~\ref{#1}}  
\newcommand{\journ}[4]{#1 #2 (#3) #4}
\newcommand{\PL}[3]{\journ{Phys.\ Lett.}{#1}{#2}{#3}}
\newcommand{\NP}[3]{\journ{Nucl.\ Phys.}{#1}{#2}{#3}}
\newcommand{\PR}[3]{\journ{Phys.\ Rev.}{#1}{#2}{#3}}
\newcommand{\PRL}[3]{\journ{Phys.\ Rev. Lett.}{#1}{#2}{#3}}
\newcommand{\ZP}[3]{\journ{Z.\ Phys.}{#1}{#2}{#3}}
\newcommand{\beq}{\begin{equation}}
\newcommand{\eeq}{\end{equation}}
\newcommand{\bea}{\begin{eqnarray}}
\newcommand{\eea}{\end{eqnarray}}

\begin{document}
\begin{frontmatter}
\title{
Next-to-Leading Order QCD Analysis of Polarized Deep
Inelastic Scattering Data}
\collab{E154 Collaboration}
\author[19]{K. Abe}, 	  	
\author[16]{T. Akagi},
\author[8]{B. D. Anderson},
\author[16]{P. L. Anthony},
\author[1]{R. G. Arnold},
\author[6]{T. Averett},
\author[21]{H. R. Band},
\author[9]{C. M. Berisso},
\author[14]{P. Bogorad},
\author[7]{H. Borel},
\author[1]{P. E. Bosted},
\author[2]{V. Breton},
\author[16]{M. J. Buenerd\thanksref{AA}},
\author[14]{G. D. Cates},
\author[10]{T. E. Chupp},
\author[9]{S. Churchwell},
\author[10]{K. P. Coulter},
\author[16]{M. Daoudi},
\author[15]{P. Decowski},
\author[16]{R. Erickson},
\author[1]{J. N. Fellbaum},
\author[2]{H. Fonvieille},
\author[16]{R. Gearhart},
\author[5]{V. Ghazikhanian},
\author[20]{K. A. Griffioen},
\author[9]{R. S. Hicks},
\author[17]{R. Holmes},
\author[6]{E. W. Hughes},
\author[5]{G. Igo},
\author[2]{S. Incerti},
\author[21]{J. R. Johnson},
\author[17]{W. Kahl},
\author[8]{M. Khayat},
\author[9]{Yu. G. Kolomensky},
\author[12]{S. E. Kuhn},
\author[14]{K. Kumar},
\author[19]{M. Kuriki},
\author[7]{R. Lombard-Nelsen},
\author[8]{D. M. Manley},
\author[7]{J. Marroncle},
\author[16]{T. Maruyama},
\author[15.5]{T. Marvin},
\author[16]{W. Meyer\thanksref{BB}},
\author[18]{Z.-E. Meziani},
\author[11.5]{D. Miller},
\author[21]{G. Mitchell},
\author[8]{M. Olson},
\author[9]{G. A. Peterson},
\author[8]{G. G. Petratos},
\author[16]{R. Pitthan},
\author[21]{R. Prepost},
\author[13]{P. Raines},
\author[12]{B. A. Raue\thanksref{CC}},
\author[1]{D. Reyna},
\author[16]{L. S. Rochester},
\author[1]{S. E. Rock},
\author[14]{M. V. Romalis},
\author[7]{F. Sabatie},
\author[4]{G. Shapiro},
\author[9]{J. Shaw},
\author[10]{T. B. Smith},
\author[1]{L. Sorrell},
\author[17]{P. A. Souder},
\author[7]{F. Staley},
\author[16]{S. St. Lorant},
\author[16]{L. M. Stuart},
\author[19]{F. Suekane},
\author[1]{Z. M. Szalata},
\author[7]{Y. Terrien},
\author[11]{A. K. Thompson},
\author[1]{T. Toole},
\author[17]{X. Wang},
\author[8]{J. W. Watson},
\author[10]{R. C. Welsh},
\author[12]{F. R. Wesselmann},
\author[21]{T. Wright},
\author[16]{C. C. Young},
\author[16]{B. Youngman},
\author[19]{H. Yuta},
\author[8]{W.-M. Zhang} and
\author[18]{P. Zyla}
\address[1]{The American University, Washington D.C. 20016}
\address[2]{Universit\'{e} Blaise Pascal, LPC IN2P3/CNRS, F-63170
Aubi\`{e}re Cedex, France} 
\address[4]{University of California, Berkeley,
California 94720-7300}
\address[5]{University of California, Los Angeles,
California  90024} 
\address[6]{California Institute of Technology, Pasadena, California 91125}
\address[7]{Centre d'Etudes de Saclay, DAPNIA/SPhN, F-91191
Gif-sur-Yvette, France} 
\address[8]{Kent State University, Kent, Ohio 44242}
\address[9]{University of Massachusetts, Amherst, Massachusetts 01003}
\address[10]{University of Michigan, Ann Arbor, Michigan 48109}
\address[11]{National Institute of Standards and Technology,
Gaithersburg, Maryland 20899}
\address[11.5]{Northwestern University, Evanston, Illinois 60201} 
\address[12]{Old Dominion University, Norfolk, Virginia  23529}
\address[13]{University of Pennsylvania, Philadelphia, Pennsylvania 19104}
\address[14]{Princeton University, Princeton, New Jersey 08544}
\address[15]{Smith College, Northampton, Massachusetts  01063}
\address[15.5]{Southern Oregon State College, Ashland, Oregon  97520}
\address[16]{Stanford Linear Accelerator Center, Stanford,
California 94309} 
\address[17]{Syracuse University, Syracuse, New York 13210}
\address[18]{Temple University, Philadelphia, Pennsylvania 19122}
\address[19]{Tohoku University, Aramaki Aza Aoba, Sendai, Miyagi, Japan}
\address[20]{College of William and Mary, Williamsburg, Virginia 23187}
\address[21]{University of Wisconsin, Madison, Wisconsin 53706}
\thanks[AA]{Permanent Address: Institut des Sciences Nucl\'{e}aires,
IN2P3/CNRS, 38026 Grenoble Cedex, France}
\thanks[BB]{Permanent Address: University of Bochum, D-44780 Bochum, Germany}
\thanks[CC]{Present Address: Florida International
University, Miami, FL 33199}
 
\begin{abstract}
We present a Next-to-Leading order perturbative QCD analysis of world
data on the spin dependent structure functions $g_1^p$, $g_1^n$, and
$g_1^d$, including the new experimental information on the
$Q^2$ dependence of $g_1^n$. Careful attention is paid to the
experimental and theoretical uncertainties. The data constrain the
first moments of the polarized valence quark distributions, but only
qualitatively constrain the polarized sea quark and gluon
distributions. The NLO results are used to determine the $Q^2$
dependence of the ratio $g_1/F_1$ and evolve the experimental data to
a constant $Q^2=5~{\rm GeV}^2$.  We determine the first moments of the
polarized structure functions of the proton and neutron and find
agreement with the Bjorken sum rule.
\end{abstract}
\end{frontmatter}

\newpage

\section{Introduction}

Next-to-Leading Order (NLO) perturbative QCD (pQCD) analyses of
unpolarized lepton-nucleon deep inelastic scattering (DIS)
\cite{ref:MRS96,ref:GRV94,ref:CTEQ} have resulted in the decomposition
of the structure functions into valence quarks, sea quarks (of each
flavor), and gluons.  The data upon which these analyses are based
include the scattering of charged leptons, neutrinos, and
antineutrinos off a variety of targets, including both protons and
deuterons, over a large kinematic range in both Bjorken $x$ and
momentum transfer $Q^2$.

Presently data are also available for polarized DIS
\cite{E80,E130,EMC,E142,E143p,E143d,E143Q2,SMCp,SMCd,E154,HERMES}.
Values for the polarized structure functions $g_1(x)$ have been
measured for protons, neutrons, and deuterons over a reasonable region
of $x$ and $Q^2$ with good precision.  Analyses of the first moments
of the structure functions, $\Gamma_1=\int g_1(x)dx$, have indicated
that relatively little of the spin of the nucleon is carried by the
quarks, suggesting that perhaps the sea quarks and gluons are
polarized.  Hence it is desirable to decompose the
spin-dependent structure functions into contributions from valence
quarks, antiquarks, and gluons just as has been done for the
spin-averaged structure functions.

On the theoretical side, a full calculation of the Next-to-Leading
Order (NLO) spin-dependent anomalous dimensions has been recently
completed~\cite{ref:NLOsplit}. This provides for a perturbative QCD
analysis of polarized DIS as a tool for decomposing the structure
functions~\cite{ref:GRSV,ref:bfr,ref:GS,ref:abfr}.  However, the lack of
polarized neutrino data and the limited kinematic coverage in $x$ and
$Q^2$ of the polarized DIS data limits the conclusions that can be
drawn.

We have recently reported on a precision measurement of the neutron
spin-dependent structure function $g_1^n$ at an average four-momentum
transfer squared $Q^2 = 5$ GeV$^2$ in SLAC experiment
E154~\cite{E154}. The two independent spectrometers used in E154
provided a possibility of studying the $Q^2$ dependence of the
structure function $g_1^n$, and extended the kinematic range of
the measurement beyond that of the previous SLAC
experiments~\cite{E142,E143p,E143d,E143Q2} to $0.014 \le x \le 0.7$
and $1~{\rm GeV}^2 \le Q^2 \le 17~{\rm GeV}^2$. The E154 results
presented in this Letter supplement our previously published
data~\cite{E154}. They currently constitute the most precise
determination of $g_1^n$. 

Of special interest for our data is the observation that the absolute
value of $g_1^n$ increases rapidly as $x$ becomes small for $x<0.1$,
approximately as $x^{-0.9}$~\cite{E154}.  This is in
striking contrast with the assumption of Regge behavior, which
suggests that $g_1^n$ is constant or decreases in magnitude with
decreasing $x$~\cite{Regge}. Moreover, if the observed $x$-dependence
of $g_1^n$ persists to $x=0$, the first moment $\Gamma_1^n$ becomes
unrealistically large.

We will show that by using NLO pQCD and reasonable assumptions about
the relation of the polarized and unpolarized distributions, we can
obtain excellent fits to our data which can be used to determine the
first moments $\Gamma_1^p$ and $\Gamma_1^n$.  Based on these fits, we
evaluate what we know about the polarization of gluons and sea
quarks.  Careful attention is paid to the theoretical and experimental
errors involved in the analysis.

\section{Formalism}

In the QCD-improved quark-parton model (QPM), the polarized structure
function $g_1(x)$ of the nucleon is related to the polarized quark,
antiquark, and gluon distributions $\Delta q(x)$, $\Delta \bar{q}(x)$,
and $\Delta G(x)$ via the factorization theorem
\cite{ref:factorization}
\beq
g_1(x,Q^2) = \frac{1}{2} \sum_q^{N_f} e^2_q 
\left[ C_q \otimes \left( \Delta q + \Delta \bar{q} \right) +
\frac{1}{N_f} C_G \otimes \Delta G \right]
\label{eq:g1NLO_c5}
\eeq
with the convolution $\otimes$ defined as 
\beq
\left( C \otimes q \right)(x,Q^2) = \int_x^1 \frac{dz}{z} 
C \left( \frac{x}{z},\alpha_S \right) q(z,Q^2).
\eeq
The sum is over all active quark flavors $N_f$. 

The first moments of the structure functions of the proton and
neutron, $\Gamma_1^p$ and $\Gamma_1^n$, allow one to test the fundamental
Bjorken sum rule~\cite{BjSum} and determine the helicity content of
the proton.  The information on the $x$ and $Q^2$ dependence gives
insight into the perturbative and non-perturbative dynamics of quarks
and gluons inside the nucleon.  Coefficient functions
$C_{q,G}(x,\alpha_S)$ correspond to the hard scattering photon-quark
and photon-gluon cross sections and are referred to as Wilson
coefficients. They are calculated in pQCD as an expansion in powers of
the strong coupling constant $\alpha_S$.  In leading order, $C^{(0)}_q
= \delta(1-x)$ and $C^{(0)}_G = 0$ according to the simple partonic
picture.  The polarized NLO coefficient functions $C^{(1)}_q$ and
$C^{(1)}_G$ in the modified minimal subtraction (\MSbar)
renormalization and factorization schemes are given in
\citeref{ref:NLOsplit}.  Throughout this paper, we use the
fixed-flavor scheme \cite{ref:GRV94,ref:GRSV} and set $N_f=3$ in
Eq.~(\ref{eq:g1NLO_c5}).  The heavy quark contributions are included
in the running of the strong coupling constant $\alpha_S(Q^2)$ calculated
to two loops~\cite{RPP96}.  For consistency with the evolution of the
unpolarized distributions, we adopt the values of $\alpha_S(Q^2)$ and
current quark masses from Ref.~\cite{ref:GRV94} that correspond to
$\alpha_S(M_Z^2) = 0.109$ or $\alpha_S(5~\mbox{GeV}^2) = 0.237$.  We
include the uncertainty associated with the value of $\alpha_S$ as
will be discussed below. The parton distributions in
Eq.~(\ref{eq:g1NLO_c5}) are those of the proton. The neutron structure
function is obtained by the isospin interchange $u \Leftrightarrow d$,
and the deuteron structure function is defined as
$g_1^d = (1/2)(g_1^p+g_1^n)(1-1.5\omega_D)$,
where the $D$-state probability
$\omega_D = 0.05 \pm 0.01$~\cite{ref:omegaD}.

The $Q^2$ evolution of the parton densities is governed by the DGLAP
equations \cite{ref:DGLAP}
\bea
Q^2\frac{d}{dQ^2}\Delta q_{\rm NS}^{\eta}(x,Q^2) & = & 
\frac{\alpha_S(Q^2)}{2\pi} P_{\rm NS}^{\eta} \otimes \Delta q_{\rm NS}^{\eta},
\quad \eta = \pm 1  \nonumber \\
\mbox{and} \quad Q^2\frac{d}{dQ^2} 
\pmatrix{\Delta\Sigma(x,Q^2)\cr \Delta G(x,Q^2)\cr} &=&
\frac{\alpha_S(Q^2)}{2\pi}
\pmatrix{P_{qq}& P_{qG}\cr P_{Gq}& P_{GG}\cr}
\otimes \pmatrix{\Delta\Sigma\cr \Delta G\cr} \; ,
\label{eq:DGLAP} 
\eea
where the index NS stands for the the non-singlet quark distributions:
valence 
$\Delta u_V(x,Q^2) = \Delta u-\Delta \bar{u}$, 
$\Delta d_V(x,Q^2) = \Delta d-\Delta \bar{d}$,
and the SU$(3)_{\rm flavor}$ non-singlet combinations 
$\Delta q_3(x,Q^2) = (\Delta u+\Delta\bar{u})-(\Delta d+\Delta\bar{d})$ and
$\Delta q_8(x,Q^2) = (\Delta u+\Delta\bar{u})+(\Delta d+\Delta\bar{d})
 - 2(\Delta s+\Delta\bar{s})$. 
The SU$(3)_{\rm flavor}$ singlet distribution is 
$\Delta\Sigma(x,Q^2) = (\Delta u+\Delta\bar{u}) + (\Delta d+\Delta\bar{d}) 
+(\Delta s+\Delta\bar{s})$. 
The index $\eta=1$ refers to the evolution of the valence
(charge-conjugation odd) distributions $u_V$ and $d_V$, and $\eta=-1$
refers to the evolution of the charge-conjugation even combinations
$\Delta q_3$, $\Delta q_8$, and $\Delta\Sigma$.
The splitting functions $P_{\rm NS}^{\eta}$ and $P_{ij}$ are
calculated perturbatively
with the leading order functions given in \citeref{ref:DGLAP}, and
the next-to-leading order expressions recently obtained in
\citeref{ref:NLOsplit}. In leading order, the evolution
of both types of non-singlet  distributions  is the same: 
$P_{\rm NS}^{(0)\eta=-1} = P_{\rm NS}^{(0)\eta=+1} = P_{qq}^{(0)}$ 
and the differences only appear in next-to-leading order.
Starting with a parametrization of the parton densities at low
initial scale $Q^2_0=0.34~{\rm GeV}^2$, the distributions at any value
of $Q^2 > Q^2_0$ are obtained using the solutions of the NLO DGLAP
equations in the Mellin $n$-moment space~\cite{ref:FurmPetr,ref:GRV90}.
The structure functions evolved in Mellin space are inverted back to
Bjorken $x$ space using the prescription of \citeref{ref:GRV90}.

One of the conventions relevant to the
interpretation of the deep
inelastic scattering data at next-to-leading order arises form the
relative freedom in defining the hard scattering cross sections
$C^{(1)}_{q,G}$ and the singlet quark density $\Delta \Sigma$ in
\eqref{eq:g1NLO_c5}, known as the factorization scheme
dependence~\cite{ref:FurmPetr,ref:Bodwin,ref:chyla}. In
the unpolarized case, the factorization scheme is fixed by specifying the
renormalization procedure for the hard scattering cross sections
$C_{q,G}$~\cite{ref:FurmPetr,ref:chyla}. In the polarized
case, the situation is further complicated by the freedom in the 
definition of the $\gamma_5$ matrix and the Levi-Civita tensor in $n
\ne 4$ dimensions~\cite{ref:Bodwin,ref:Manohar} in dimensional
regularization~\cite{ref:DimReg}.  
The NLO splitting functions and Wilson coefficients are
given in \citeref{ref:NLOsplit} in the \MSbar\ scheme
with the definition of the $\gamma_5$ matrix following
\citeref{ref:DimReg}. The specific feature of this scheme is that the
first moment of the gluon coefficient function vanishes
$C_G^{(1)}(n=1) = 0$, and the gluon density does not contribute to the
integral of $g_1$. 
Several authors~\cite{DeltagI,DeltagII,DeltagIII} have advocated using
a different factorization scheme in which the axial anomaly
contribution $-(\alpha_S(Q^2)/4\pi) \sum_q e^2_q \Delta G$ is included
in the integral $\Gamma_1$. The suggestion generated a vivid
theoretical
debate~\cite{ref:Bodwin,DeltagI,DeltagII,DeltagIII,ref:JaffeManohar}.
Such a scheme was referred to in \citeref{ref:bfr} as the
Adler--Bardeen (AB) scheme.  In the AB scheme the total quark helicity
is redefined compared to the \MSbar\ scheme
\bea
\Delta \Sigma_{\rm AB} & = & \Delta q_0(Q^2) + 
\frac{N_f \alpha_S(Q^2)}{2\pi} \Delta G(Q^2) \; , \nonumber \\
\Delta\Sigma_{\overline{\rm MS}} & = & \Delta q_0(Q^2) \; ,
\label{eq:anomaly_c5}
\eea
where $\Delta q_0$ is the proton matrix element of the SU$(3)_{\rm
flavor}$ singlet axial current.  An attractive feature of the AB
scheme is that $\Delta\Sigma_{\rm AB}$ is independent of $Q^2$ even
beyond the leading order. One could also resurrect the naive QPM
expectation $\Delta \Sigma \approx 0.6 - 0.7$ and explain the
violation of the Ellis-Jaffe sum rule if the product
$\alpha_S(Q^2)\Delta G(Q^2)$ turned out to be
large~\cite{DeltagI,DeltagII,DeltagIII}.

The product $\alpha_S(Q^2)\Delta G(Q^2)$ is scale-independent in
the leading order since its anomalous dimension expansion starts at
order $\alpha_S^2$ \cite{ref:Kod80}. This implies that as $\alpha_S$
decreases logarithmically with $Q^2$, $\Delta G$ grows as
$1/\alpha_S(Q^2)$. This growth is compensated by the increasing (with
opposite sign) orbital angular momentum contribution $\langle L_z
\rangle$~\cite{ref:AEL,ref:DeltaG_Lz} in order to satisfy the 
sum rule
\begin{equation}
\frac{1}{2}\Delta \Sigma + \Delta G + \langle L_z \rangle = \frac{1}{2}.
\end{equation}
The gauge-invariant and scheme-independent formulation of this sum
rule has recently been presented in \citeref{ref:Jigauge}. 

Another consequence is that the ambiguity in the definition of the
total quark helicity in \eqref{eq:anomaly_c5} does not vanish at
infinite $Q^2$. However, as long as the factorization and
renormalization schemes are used consistently, NLO predictions can be
made for the spin dependent structure functions and other hadronic
processes involving spin degrees of freedom once the parton
distributions are determined in one scheme and at one scale.

A transformation from the \MSbar\ scheme of t'Hooft and
Veltman~\cite{ref:DimReg} to the AB
scheme was constructed in \citeref{ref:bfr}. 
This scheme is a simple modification of \MSbar\
since it preserves the low and high $x$ behavior of the coefficient
functions and anomalous dimensions, and thus the asymptotic behavior of
parton distributions is not modified. In order to demonstrate the
effects of the factorization scheme dependence, we perform our
calculations in both $\overline{\mbox{MS}}$ and AB schemes.

\section{Fits}

Following \citeref{ref:GRSV}, we make our central ansatz of
parametrizing the polarized parton distribution at the low initial
scale $Q^2_0 = 0.34$~GeV$^2$ as follows:
\begin{equation}
\Delta f(x,Q^2_0) = A_f x^{\alpha_f} (1-x)^{\beta_f} f(x,Q^2_0) \; ,
\label{eq:parametrization} 
\end{equation}
where 
$\Delta f=\Delta u_V,\: \Delta d_V,\: \Delta \bar{Q}, \: \Delta G$ 
are the polarized valence, sea, and gluon distributions (see below for
the definition of $\Delta\bar{Q}$),
and $f(x,Q^2_0)$ are the unpolarized parton distributions from
\citeref{ref:GRV94}. The parametrization assumes the power-like
asymptotic behavior of the polarized distributions at low $x$ and low
$Q^2$, namely $\Delta f \sim x^{\gamma_f},\: x \rightarrow 0$, where
$\gamma_f$ is the sum of the polarized power $\alpha_f$ and the low
$x$ power of the unpolarized distribution. 
Since inclusive deep inelastic scattering does not provide
sufficient information about the flavor separation of the polarized
sea, we assume an ``isospin-symmetric'' sea
$
\Delta\bar{u} = \Delta\bar{d}\equiv 
\frac{1}{2}\left(\Delta\bar{u} + \Delta\bar{d}\right)
$.
Under this assumption, the sea quark contribution to the polarized
structure functions of the proton and neutron is the same: 
\beq
g_1^{p\ {\rm sea}} = g_1^{n\ {\rm sea}} = 
(5/9)C_q \otimes \left[
1/2(\Delta\bar{u}+\Delta\bar{d})+1/5\Delta\bar{s} \right] \; .
\label{eq:g1sea}
\eeq
Inclusive DIS does not probe the light and strange sea
independently. The only sensitivity to the
difference between $\Delta\bar{u}$, $\Delta\bar{d}$, and
$\Delta\bar{s}$ comes from the difference in the evolution of the two
types of non-singlet distributions ($\eta=\pm 1$ in
\eqref{eq:DGLAP}). Such a difference is beyond the reach of 
present-day experiments. Hence, we will parametrize a particular
combination of the sea quark distributions that appears in
\eqref{eq:g1sea}: 
\beq
\Delta \bar{Q} \equiv 1/2(\Delta \bar{u} + \Delta \bar{d}) + 
1/5 \Delta \bar{s} \;  .
\label{eq:barQ}
\eeq 
Furthermore, we assume the $x$ dependence of the
polarized strange and light sea to be the same, and fix the
normalization of the strange sea by 
\begin{equation}
\Delta s = \Delta\bar{s}=\lambda_s\frac{\Delta \bar{u} +\Delta  \bar{d}}{2} = 
\frac{\lambda_s}{1+\lambda_s/5}\Delta \bar{Q} \: ,
\end{equation}
with the SU$(3)_{\rm flavor}$ symmetry breaking parameter $\lambda_s$
varying between $1$ and $0$ (where the latter choice corresponds to
an unpolarized strange sea). 

The positivity constraint,
$
|\Delta f(x)| \le f(x)
$,
satisfied (within uncertainties) at the initial scale $Q^2_0$, holds
at all scales $Q^2 > Q^2_0$; it leads to the constraints $\alpha_f \ge
0$ and $\beta_f \ge 0$.  In addition, we assume the helicity retention
properties of the parton distributions \cite{ref:bbs} that
require\footnote{We have checked that the data are consistent with
this assumption.}  $\beta_f = 0$.  Unlike most NLO
analyses~\cite{ref:GRSV,ref:bfr,ref:GS}, we do not assume SU$(3)_{\rm
flavor}$ symmetry and do not fix the normalization of the non-singlet
distributions by the axial charges $\Delta q_3=F+D$ and $\Delta
q_8=3F-D$, where $F$ and $D$ are the antisymmetric and symmetric
SU$(3)$ coupling constants of hyperon beta decays~\cite{Bour}. Thus,
we are able to test the Bjorken sum rule. In addition, the structure
functions are not sensitive to the corrections beyond NLO in the
data range.

The remaining eight coefficients are determined by fitting the
available data on the spin dependent structure functions $g_1^{p,n,d}$
of the proton~\cite{EMC,E143p,E143Q2,SMCp},
neutron~\cite{E142,E154,HERMES}, and deuteron~\cite{E143d,E143Q2,SMCd}
with $Q^2 > 1$~GeV$^2$. We use either the results for $g_1$ or
determine the structure functions at the experimental values of $Q^2$
using the results for $g_1/F_1$~\cite{ref:formulae}. The
unpolarized structure function $F_1$ is obtained from a recent
parametrization of $F_2(x,Q^2)$ from NMC~\cite{ref:NMC} and a fit
to the data on $R(x,Q^2)$, the ratio of longitudinal to transverse
photoabsorption cross sections from SLAC~\cite{ref:Whitlow}. The
weight of each point is determined by the statistical error.  The best
fit coefficients are listed in \tabref{tab:coeff}. The total $\chi^2$
of the fits are $146$ and $148$ for $168$ points in \MSbar\ and AB
schemes, respectively.

The statistical errors on extracted parton densities $\Delta
q(x,Q^2)$, $\Delta \bar{q}(x,Q^2)$, and $\Delta G(x,Q^2)$ were
calculated by adding in quadrature statistical contributions from
experimental points. The weight of each point was obtained by varying
the point within its statistical error and calculating the change in
the parton density~\cite{ref:YuryThesis}. The systematic error is
usually dominated by the normalization errors (target and beam
polarizations, dilution factors, etc.). Thus the systematic errors are
to a large extent correlated point to point within one
experiment\footnote{This includes both proton and deuteron data taken
in a single experiment, such as E143 and SMC.}. We therefore assumed
100\% correlated systematic errors for any given experiment and added
systematic contributions within one experiment linearly. Systematic
errors for each experiment were then added quadratically to obtain 
systematic uncertainties on parton densities.

The biggest source of theoretical uncertainty comes from the
uncertainty on the value of $\alpha_S$. We estimate it by repeating
the fits\footnote{We also relax the positivity constraints.}  with
$\alpha_S(M_Z^2)$ varied in the range allowed by the unpolarized fixed
target DIS experiments~\cite{RPP96} $\alpha_S(M_Z^2) = 0.108-0.116$.
The quality of the fits deteriorated significantly when the values as
high as $\alpha_S(M_Z^2) = 0.120$ were used.  The scale uncertainty is
included in the error on $\alpha_S$.  We also vary current quark
masses in the range $m_c = 1 - 2$~GeV and $m_b = 4 - 5$~GeV which
affects the running of $\alpha_S$. The effect of SU$(3)_{\rm flavor}$
breaking is estimated by varying the parameter $\lambda_s$ from $1$ to
$0$. These factors are found to have a small influence on the results.
To test the sensitivity to the shape of the initial distributions and
the value of the starting scale $Q^2_0$, we repeat the fit with
initial unpolarized distributions taken from \citeref{ref:MRS96} at
$Q^2_0 = 1$~GeV$^2$ and find the results consistent with values given
in Tables \ref{tab:coeff} and \ref{tab:moments} within quoted
statistical uncertainties. Possible higher twist effects are neglected
since they are expected to drop with the photon-nucleon invariant mass
squared $W^2$ as $1/W^2$~\cite{ref:bbs19}. The cut $W^2 > 4$~GeV$^2$
has been applied to all the data with the majority of them exceeding
$W^2 > 8$~GeV$^2$.

\section{Results and discussion}

\begin{figure}
\begin{center}
\epsfig{file=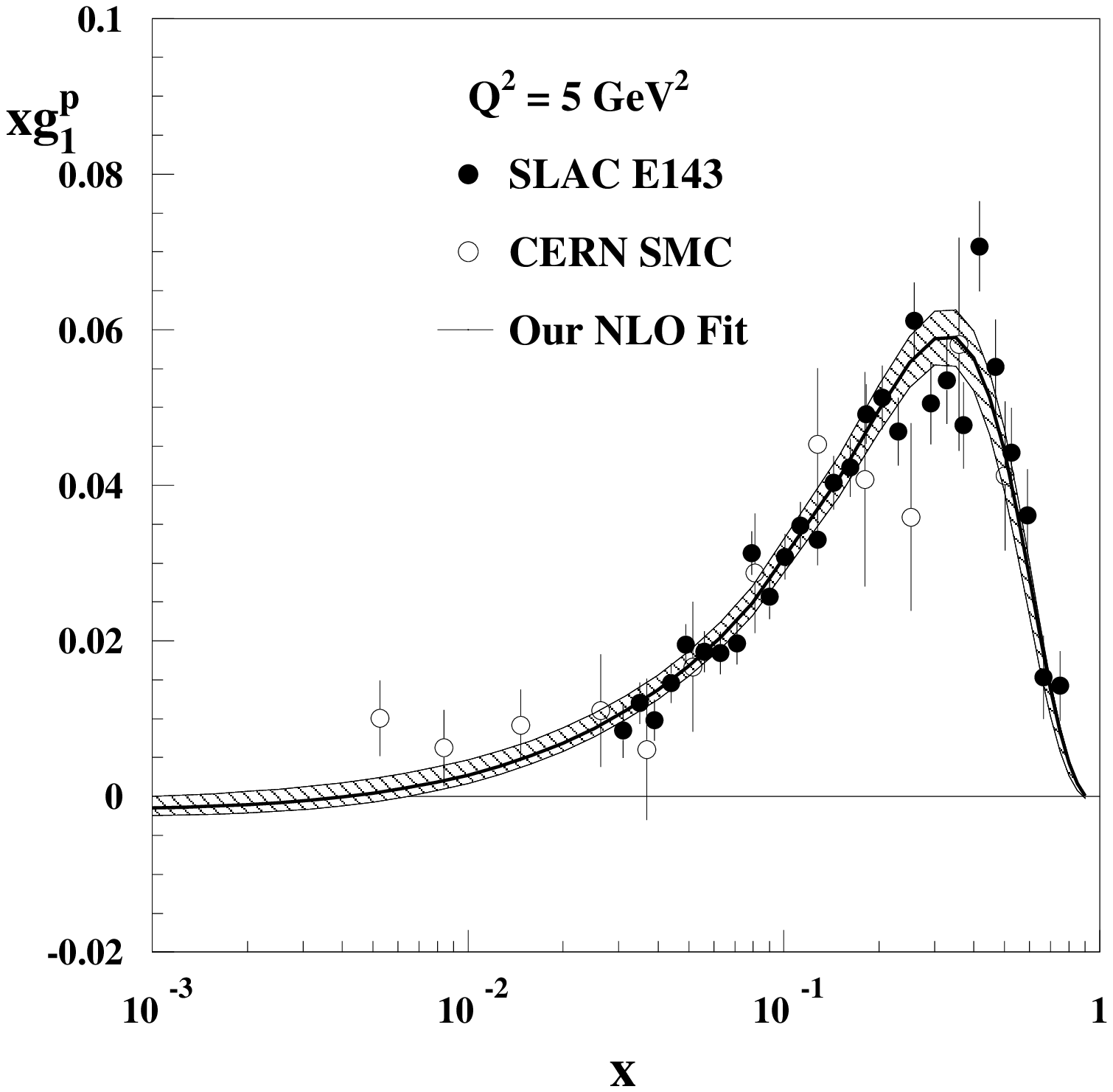,width=2.5in}
\epsfig{file=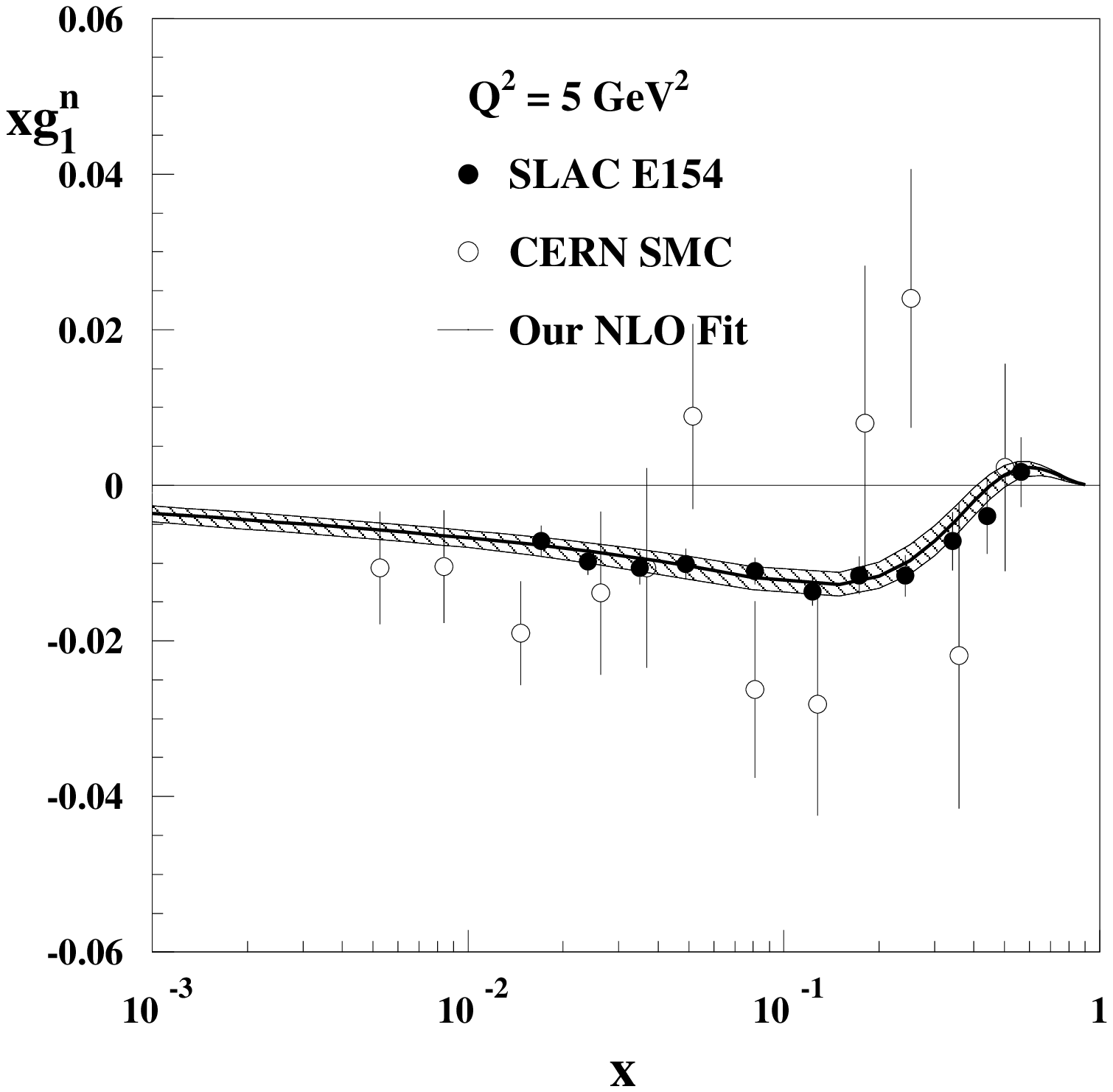,width=2.5in}
\end{center}
\caption{The structure functions $xg_1^p$ and 
$xg_1^n$ at $Q^2 = 5$~GeV$^2$.  The E143, SMC, and E154 data have been
evolved to $Q^2 = 5$~GeV$^2$ using a procedure described in the
text. The result of the $\overline{\mbox{MS}}$ fit is shown by the
solid line and the hatched area represents the total error of the
fit.}
\label{fig:g1_5}
\end{figure}

Results for the structure functions of the proton and neutron $g_1^p$
and $g_1^n$ at 5~GeV$^2$ are compared to the experimental data in
\figref{fig:g1_5}. Despite a small number of free parameters, the
fits are excellent.  In addition, at the initial scale $Q^2=0.34~{\rm
GeV}^2$ the low $x$ behavior of the distributions is consistent with
the Regge theory prediction $\gamma_f\approx 0$~\cite{Regge}.
However, Regge theory in the past has been applied at the $Q^2\sim
2-10~{\rm GeV}^2$ of the experiments. This procedure clearly cannot be
applied to the E154 neutron data for $0.014<x<0.1$, and is
incompatible with the pQCD
predictions~\cite{ref:nlo_lowx,ref:lowx_bfr}.  If instead, Regge
theory is assumed to apply at our starting $Q^2$ to fix the values of
powers $\gamma_f$ to anywhere between 0 and 1, the data are fit with
four parameters.  Two of those parameters control the small
contributions of gluons and antiquarks. When these parameters are
fixed to zero, the resulting fit {\it with only two free parameters}
still provides a reasonable description of the data everywhere except
the low $x$ region where it underestimates the E154 data on $g_1^n$ by
about two standard deviations.

The values of the first moments of parton distributions, as well as
the first moments of structure functions at $Q^2=5$~GeV$^2$, are given
in \tabref{tab:moments}.  The procedure of fitting structure functions
to power laws at low $Q^2=0.34~{\rm GeV}^2$ evolved up to the
experimental $Q^2$ results in low $x$ behavior that can be integrated
to yield the first moments.  If instead, the data are fit to a power
law in $x$ at the average $Q^2\approx 5~{\rm GeV}^2$, a significantly
bigger first moment of $g_1^n$ is obtained~\cite{ref:YuryThesis}.
Hence our results for the first moments depend strongly on the
assumptions that we make regarding the low $x$ behavior.  However, the
simple assumptions that we made are attractive theoretically and have
remarkable predictive power.

The first moment of the deuteron structure function $g_1^d$ that we
obtain is smaller than that of \citeref{E143d}.  The reason is that
our assumptions about the low $x$ behavior of $g_1$ result in a
contribution beyond the measured region of 
$\int_{0}^{0.03} g_1^d\ dx \approx-0.014$ 
as opposed to $\approx +0.001$ estimated in
\citeref{E143d} assuming Regge behavior at $Q^2=3~{\rm GeV}^2$.  The
first moment of $g_1^p$ is numerically less sensitive to how the data
are extrapolated.

The first moments of the valence quark distributions are determined
well, but the moments of the sea quark and gluon distributions are
only qualitatively constrained.  We note that the contribution of the
experimental systematic errors to the errors on the first moments of
the parton distributions is comparable to the statistical
contribution. The full error on the first moment of the gluon
distribution $\Delta G$ is bigger than quoted in \citeref{ref:bfr}
despite the fact that the new data from E154 were added.  The
theoretical uncertainty is also quite large; it could potentially be
reduced if the simultaneous analysis of the unpolarized and polarized
data was performed (including $\alpha_S$ as one of the parameters).
It is interesting to note that at $Q^2=0.34~{\rm GeV}^2$ the orbital
angular momentum contribution $\langle L_z\rangle=-0.2^{+0.7}_{-0.3}$
is consistent with zero, \ie\ helicities of quarks and gluons account
for most of the nucleon spin.  The results of the fits in both \MSbar\
and AB schemes are consistent within errors. The fits are
significantly less stable in the AB scheme.  Note that the values of
the singlet axial charge $\Delta q_0$ are essentially the same for
fits in both schemes.

\begin{figure}
\begin{center}
\epsfig{file=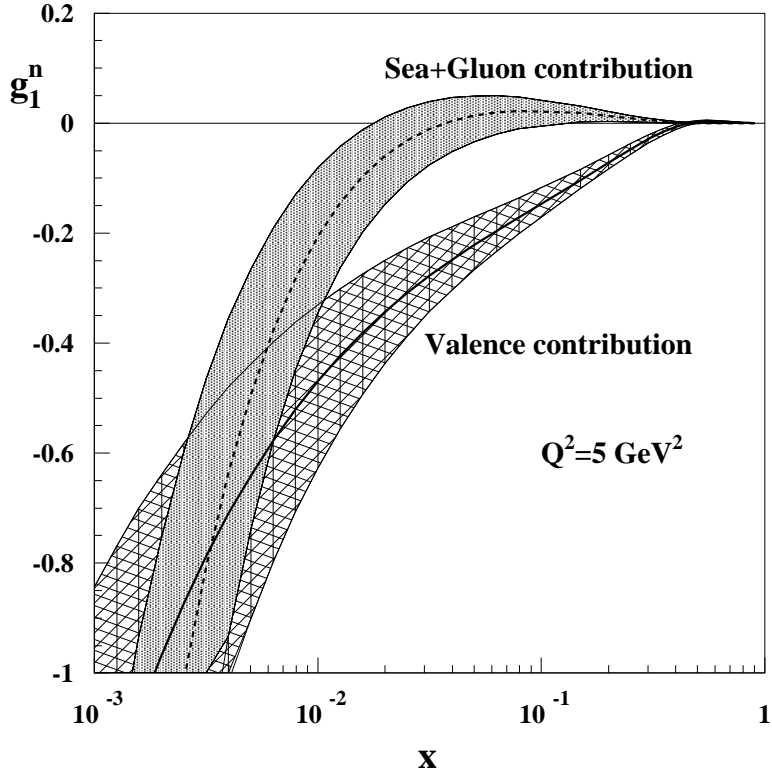,height=4in}
\end{center}
\caption{The contributions to the structure function $g_1^n$ of the
neutron from the valence quarks
[$(1/18)C_q\otimes(\Delta u_V+4\Delta d_V)$] (solid line)
and from the sea quarks and gluons 
[$(5/9)C_q\otimes\Delta\bar{Q}+(1/9)C_G\otimes\Delta G$] (dashed line). 
The shaded and hatched areas represent the total uncertainties on each
quantity.}
\label{fig:part_err}
\end{figure}

The contributions from the valence quarks 
$g_1^{n\ {\rm valence}}=(1/18)C_q\otimes(\Delta u_V+4\Delta d_V)$
and sea quarks and gluons
$g_1^{n\ {\rm sea+gluon}}=
(5/9)C_q\otimes\Delta\bar{Q}+(1/9)C_G\otimes\Delta G$
to the neutron spin structure function at $Q^2=5~{\rm GeV}^2$ are
shown in \figref{fig:part_err}. One can see that the sea and gluon
contributions are larger than the valence contributions at
$x\approx 10^{-3}$. Although the sea contributions to $g_1^n$ are
relatively modest in the E154 data range $x>0.01$, the strong $x$
dependence $g_1^n\sim x^{-0.8}$ observed by E154 below $x=0.1$ is
largely due the sea and gluon contributions.  An observation of
a negative value of $g_1^p$ at lower $x$ and higher $Q^2$ 
would provide direct evidence of a polarized sea.

One may note an apparent $\approx 2\sigma$ disagreement of $\Delta
q_3$ with the value extracted from the neutron beta-decay~\cite{RPP96}
$\Delta q_3 = g_A = 1.2601 \pm 0.0025$.  This is due to the fact that
the calculation is done in NLO, and the higher order corrections to
the Bjorken sum rule are not taken into account. The corrections can
be as large as 5\% \cite{GorLar} at $Q^2 \approx 5$~GeV$^2$. They
would bring $\Delta q_3$ in better agreement with the beta decay
data. For consistency with the NLO approximation, we do not include
this correction; it has no effect on the physical observable $g_1$.

Using the parametrization of the parton distributions, one can obtain
the polarized structure function (Eq.~(\ref{eq:g1NLO_c5})) and evolve
the experimental data points to a common $\langle Q^2 \rangle$ using
the formula: 
\begin{equation}
g_1^{\rm exp}(x_i,\langle Q^2 \rangle) = 
g_1^{\rm exp}(x_i,Q^2_i) - \Delta g_1^{\rm fit}(x_i,Q^2_i,\langle Q^2 
\rangle)  
\label{eq:g1evolved}
\end{equation}
with
\begin{equation}
\Delta g_1^{\rm fit}(x_i,Q^2_i,\langle Q^2 \rangle) = 
g_1^{\rm fit}(x_i,Q^2_i) - 
g_1^{\rm fit}(x_i,\langle Q^2 \rangle)
 \; ,
\label{eq:dg1}
\end{equation}
where $g_1^{\rm exp}(x_i,Q^2_i)$ is the structure function measured at
the experimental kinematics, and $g_1^{\rm fit}$ is the fitted
value. The errors on $g_1^{\rm exp}(x_i,\langle Q^2 \rangle)$
have three sources:
\begin{equation}
\sigma^2(g_1^{\rm exp}(x_i,\langle Q^2 \rangle)) = 
\sigma^2(g_1^{\rm exp})_{\rm stat.} + 
\sigma^2(g_1^{\rm exp})_{\rm syst.} +
\sigma^2(g_1)_{\rm evol.} \; ,
\end{equation}
where statistical and systematic uncertainties take into
account the correlation between $g_1^{\rm exp}(x_i,Q^2_i)$ and
$g_1^{\rm fit}$, and the evolution uncertainty includes only
uncorrelated experimental uncertainties as well as theoretical
uncertainties added in quadrature.  

\begin{figure}
\begin{center}
\epsfig{file=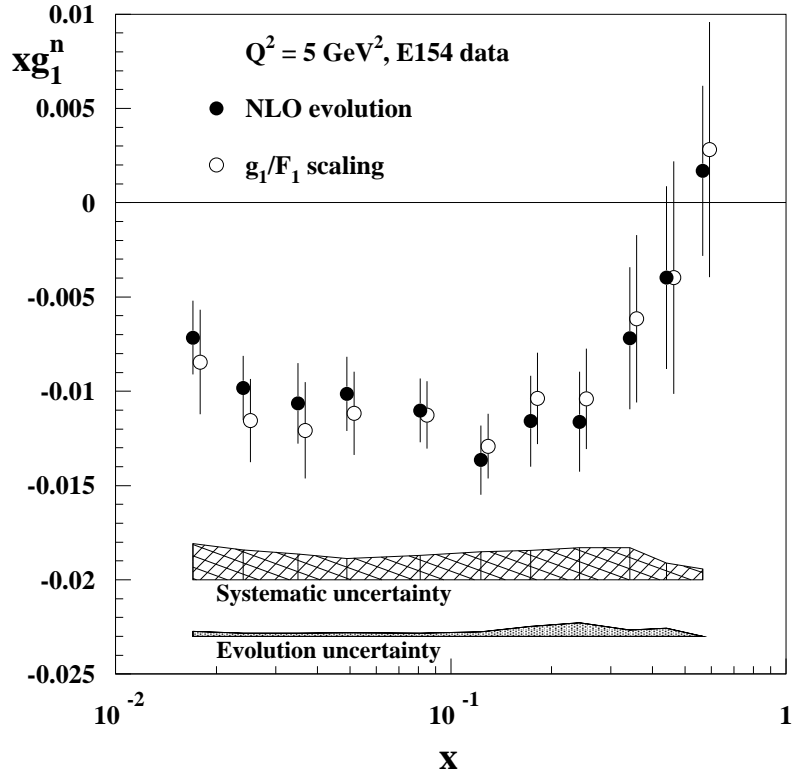,height=4in}
\end{center}
\caption{The structure function $xg_1^n$ evolved to $Q^2=5~{\rm GeV}^2$
using our \MSbar\ parametrization and using the assumption that
$g_1^n/F_1^n$ is independent of $Q^2$.}
\label{fig:evolve_e154}
\end{figure}

\begin{figure}
\begin{center}
\epsfig{file=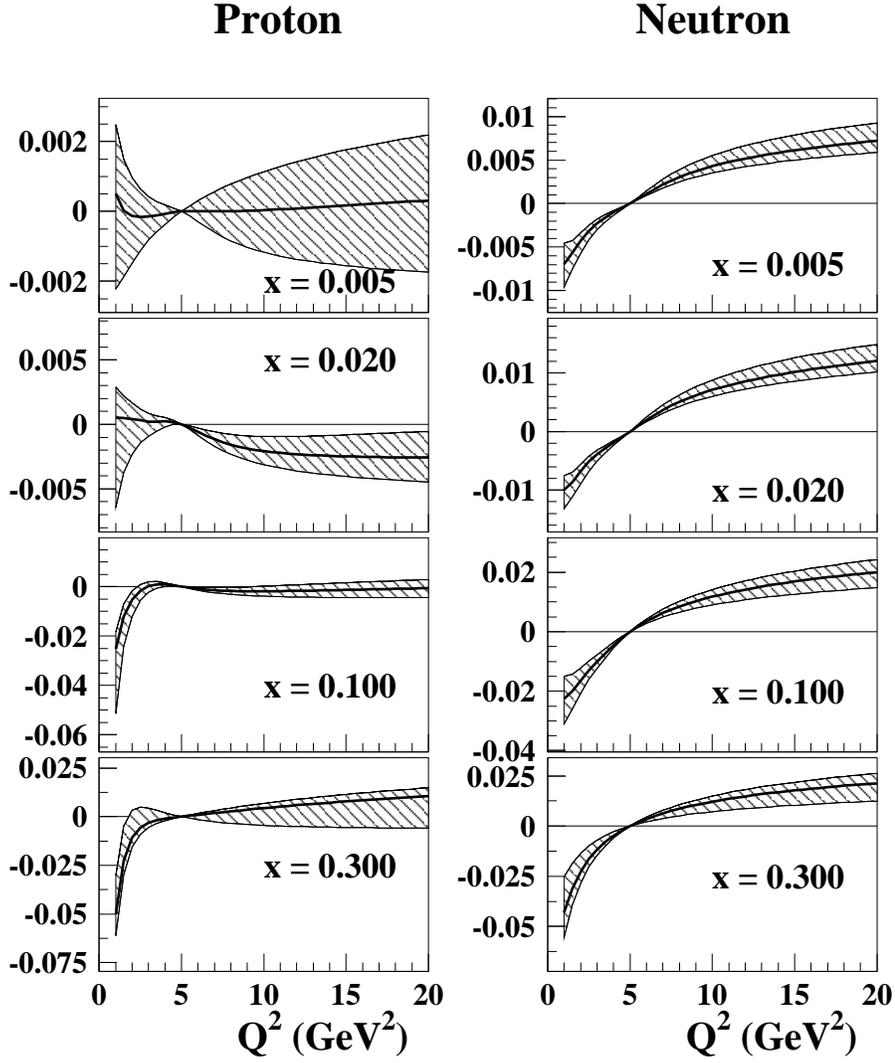,height=5.5in}
\end{center}
\caption{
Evolution of the ratios $g_1/F_1$ for proton (left) and neutron
(right). Plotted is the difference 
$\frac{g_1}{F_1}(x,Q^2)-\frac{g_1}{F_1}(x,5~{\rm
GeV}^2)$.  The \MSbar\ fit is shown by the solid line and the hatched
area represents the total (experimental and theoretical) uncertainty
of the fit.}
\label{fig:q2_nlo}
\end{figure}

The data on the structure function $g_1^n$ from two independent
spectrometers used in E154 are given in \tabref{tab:g1spec}. These
results provide new information on the $Q^2$ dependence of $g_1^n$ and
thus supplement our previously published data~\cite{E154}.
Table~\ref{tab:g1spec} also lists the E154 data points evolved to
$\langle Q^2 \rangle = 5$~GeV$^2$ using the \MSbar\ parametrization.
The NLO evolution is compared to the traditional assumption of scaling
of $g_1^n/F_1^n$ in \figref{fig:evolve_e154}. The difference is only
slightly smaller than the precision of the present-day experiments.
The effect on $\Gamma_1^n$ is small only if the integral is evaluated
at the average value of $Q^2$ (as is usually done).  The $Q^2$
dependence of the ratio $g_1/F_1$ is shown in \figref{fig:q2_nlo}. We
plot the difference between the values of $g_1/F_1$ at a given $Q^2$
and $Q^2=5~{\rm GeV}^2$ to which the SLAC data are evolved.  For the
neutron, the evolution of $g_1^n$ is slower than that of $F_1^n$.
Therefore, assuming scaling of $g_1^n/F_1^n$, one typically
overestimates the absolute value of $g_1^n(x,\langle Q^2 \rangle)$ at
low $x$ (where $Q^2_i<\langle Q^2 \rangle$), and underestimates it at
high $x$ (where $Q^2_i>\langle Q^2 \rangle$). The two effects
approximately cancel for the integral over the measured range in the
case of E154. However, the shape of the structure function at low $x$
affects the extrapolation to $x=0$. The effect of the perturbative
evolution is qualitatively the same for the proton.

The data on $g_1^n$ averaged between two
spectrometers are given in \tabref{tab:g1comb}.  
Integrating the data in the measured range, we obtain (at
$Q^2=5~{\rm GeV}^2$)
\beq
\int_{0.014}^{0.7} dx\ g_1^n(x) =  -0.035 \pm 0.003 \pm 0.005 \pm 0.001 \; ,
\label{eq:int_g1nlo} 
\eeq
where the first error is statistical, the second is systematic, and
the third is due to the uncertainty in the evolution. This value
agrees well with the number 
$-0.036\pm0.004~({\rm stat.})\pm0.005~({\rm syst.})$\cite{E154}
obtained assuming the $Q^2$ independence of $g_1^n/F_1^n$. Using the
\MSbar\ parametrization to evaluate the contributions from the
unmeasured low and high $x$ regions, we determine the first moment
\beq
\Gamma_1^n = -0.058 \pm 0.004~({\rm stat.})
		    \pm 0.007~({\rm syst.})
		    \pm 0.007~({\rm evol.})
\label{eq:Gamma1n}
\eeq
at $Q^2=5~{\rm GeV}^2$.

The behavior of the purely non-singlet combination $(g_1^p-g_1^n)(x)$
is expected to be softer at low $x$ than its singlet counterpart
\cite{ref:2logs}. Evolving the E154 neutron and E143 proton
\cite{E143p} data to $Q^2=5~{\rm GeV}^2$ and using the \MSbar\
parametrization of \tabref{tab:coeff} to determine the contributions
from the unmeasured low and high $x$ regions, we obtain for the
Bjorken sum
\beq
\Gamma_1^{p-n}(5~{\rm GeV}^2) = \int_0^1 dx \ ( g_1^p - g_1^n) = 
                                 0.171  
                             \pm 0.005 
                             \pm 0.010
                             \pm 0.006 \; ,
\eeq
where the first error is statistical, the second is systematic, and
the third is due to the uncertainty in the evolution and low $x$
extrapolation. This value is in good agreement with the
$O(\alpha_S^3)$ \cite{GorLar} prediction $0.188$ evaluated with
$\alpha_S(M_Z^2)=0.109$, and it also agrees very well with the value
in \tabref{tab:moments} obtained by direct integration of the parton
densities. The result is fairly insensitive to the details of the
low-$x$ extrapolation which is well constrained by the data. The low
$x$ behavior in the non-singlet polarized sector is also relatively
insensitive to the higher-order corrections~\cite{ref:resum}.  On the
other hand, the low-$x$ extrapolation of the proton and neutron
integrals alone still relies on the assumption that the asymptotic
behavior of sea quark and gluon distributions can be determined from
the present data, and that the effects of higher-order resummations
are small. These assumptions, and therefore the evaluation of the
total quark helicity $\Delta \Sigma$, are on potentially weaker
grounds. Precise higher energy data on the polarized structure
functions of both proton and neutron are required to determine this
quantity.

\section{Conclusions and Outlook}

Additional high precision data from SLAC experiment E155 on the polarized
structure functions of the proton and deuteron will be important in
understanding the spin structure of the nucleon. 
New results on the proton structure function $g_1^p$ from SMC have
recently been presented~\cite{ref:SMC_Derro}.
Also, the polarized electron-proton
collider experiments proposed at HERA~\cite{HERA} would be of great
importance in unraveling the low $x$ behavior of the spin-dependent
structure function $g_1^p$. Furthermore, polarized fixed-target
experiments at the Next Linear Collider would determine the structure
functions of both the proton and the neutron over a broad kinematic range
\cite{Emlyn_NLC}, and thus compliment the HERA program.
Extrapolations based on the fits in this paper suggest that $g_1^n(x)$
will be large at low $x$ and have a significant $Q^2$ dependence.
Observing $g_1^p$ become negative at low $x$ would provide
direct evidence of a polarized sea.

In conclusion, we have performed a Next-to-Leading order QCD analysis
of the world data on polarized deep inelastic scattering.  The data
constrain the first moments of the polarized valence quark
distributions; the polarized gluon and sea quark distributions can
only be qualitatively constrained. We determine that the $Q^2$
dependence of the ratio $g_1/F_1$ for the proton and neutron is
sizable compared to present experimental uncertainties. We use the NLO
pQCD evolution to determine the first moments of the spin dependent
structure functions of the proton and neutron at $Q^2=5~{\rm GeV}^2$,
and find that the data agrees with the Bjorken sum rule within one
standard deviation.

We thank the SLAC accelerator department the successful operation of
the E154 Experiment.  We would also like to thank A.\ V.\ Manohar for
reviewing the manuscript and valuable comments and W.\ Vogelsang and
S.\ Forte for stimulating discussions.  This work was supported by the
Department of Energy; by the National Science Foundation; by the Kent
State University Research Council (GGP); by the Jeffress Memorial
Trust (KAG); by the Centre National de la Recherche Scientifique and
the Commissariat a l'Energie Atomique (French groups); and by the
Japanese Ministry of Education, Science and Culture (Tohoku).

\newpage
%
%

%
\begin{table}
\caption{Fitted values of the free parameters in
Eq.~(\protect{\ref{eq:parametrization}}) in $\overline{\mbox{MS}}$ and AB
schemes. Also quoted are the statistical, systematic, and theoretical errors.}
\label{tab:coeff}
\begin{center}
\begin{tabular}{l|rccc|rccc}
\hline\hline
 & \multicolumn{4}{c|}{$\overline{\mbox{MS}}$} & \multicolumn{4}{c}{AB} \\ \hline
           & Value  &  Stat.             & Syst.              & Theory & Value & Stat. & Syst. & Theory \\ \hline
$A_u$      & $ 0.99$ & $^{+0.08}_{-0.08}$ & $^{+0.04}_{-0.05}$ & $^{+0.97}_{-0.11}$ & $ 0.98$ & $^{+0.07}_{-0.06}$ & $^{+0.05}_{-0.07}$ & $^{+0.96}_{-0.09}$ \\
$A_d$      & $-0.78$ & $^{+0.14}_{-0.20}$ & $^{+0.05}_{-0.05}$ & $^{+0.05}_{-1.28}$ & $-0.82$ & $^{+0.06}_{-0.11}$ & $^{+0.07}_{-0.06}$ & $^{+0.31}_{-1.21}$ \\
$A_Q$      & $-0.02$ & $^{+0.03}_{-0.06}$ & $^{+0.01}_{-0.02}$ & $^{+0.01}_{-0.35}$ & $-0.04$ & $^{+0.02}_{-0.05}$ & $^{+0.01}_{-0.02}$ & $^{+0.03}_{-0.06}$ \\
$A_G$      & $1.6\z$ & $^{+1.1\z}_{-0.9\z}$ & $^{+0.6\z}_{-0.6\z}$ & $^{+0.2\z}_{-1.3\z}$ & $ 0.1\z$ & $^{+1.0\z}_{-0.3\z}$ & $^{+0.5\z}_{-0.2\z}$ & $^{+0.1\z}_{-0.6\z}$ \\
$\alpha_u$ & $ 0.63$ & $^{+0.06}_{-0.07}$ & $^{+0.04}_{-0.05}$ & $^{+0.36}_{-0.06}$ & $ 0.55$ & $^{+0.08}_{-0.06}$ & $^{+0.03}_{-0.04}$ & $^{+0.56}_{-0.05}$ \\
$\alpha_d$ & $ 0.28$ & $^{+0.15}_{-0.11}$ & $^{+0.05}_{-0.03}$ & $^{+0.75}_{-0.03}$ & $ 0.40$ & $^{+0.20}_{-0.12}$ & $^{+0.07}_{-0.13}$ & $^{+0.53}_{-0.34}$ \\
$\alpha_Q$ & $ 0.04$ & $^{+0.29}_{-0.03}$ & $^{+0.12}_{-0.03}$ & $^{+0.55}_{-0.01}$ & $ 0.00$ & $^{+0.17}_{-0.00}$ & $^{+0.17}_{-0.00}$ & $^{+0.00}_{-0.00}$ \\
$\alpha_G$ & $ 0.8\z$ & $^{+0.4\z}_{-0.5\z}$ & $^{+0.3\z}_{-0.3\z}$ & $^{+0.1\z}_{-0.6\z}$ & $ 0.0\z$ & $^{+0.7\z}_{-0.0\z}$ & $^{+1.0\z}_{-0.0\z}$ & $^{+1.0\z}_{-0.0\z}$ \\
\hline\hline
\end{tabular}
\end{center}
\end{table}
\begin{table}
\caption{First moments of the polarized parton distributions and 
structure functions of the proton, neutron, and deuteron in
$\overline{\mbox{MS}}$ and AB schemes evaluated at $Q^2 = 5$~GeV$^2$.
Errors are statistical, systematic, and theoretical.}
\label{tab:moments}
\begin{center}
\begin{tabular}{l|rccc|rccc}
\hline\hline
 & \multicolumn{4}{c}{$\overline{\mbox{MS}}$} & \multicolumn{4}{c}{AB} \\ \hline
           & Value  &  Stat.             & Syst.              & Theory & Value & Stat. & Syst. & Theory \\ \hline
$\Delta u_V$     & $ 0.69$ & $^{+0.03}_{-0.02}$ & $^{+0.05}_{-0.04}$ & $^{+0.14}_{-0.01}$ & $ 0.74$ & $^{+0.03}_{-0.02}$ & $^{+0.03}_{-0.03}$ & $^{+0.07}_{-0.01}$ \\
$\Delta d_V$     & $-0.40$ & $^{+0.03}_{-0.04}$ & $^{+0.03}_{-0.03}$ & $^{+0.07}_{-0.00}$ & $-0.33$ & $^{+0.03}_{-0.04}$ & $^{+0.03}_{-0.05}$ & $^{+0.01}_{-0.03}$ \\
$\Delta \bar{Q}$ & $-0.02$ & $^{+0.01}_{-0.02}$ & $^{+0.01}_{-0.01}$ & $^{+0.00}_{-0.03}$ & $-0.03$ & $^{+0.02}_{-0.02}$ & $^{+0.01}_{-0.01}$ & $^{+0.01}_{-0.01}$ \\
$\Delta G $      & $1.8\phantom{0}$ & $^{+0.6} _{-0.7} $ & $^{+0.4}_{-0.5}$ & $^{+0.1}_{-0.6}$  & $0.4\phantom{0}$ & $^{+1.0}_{-0.7}$ & $^{+0.9}_{-0.6}$ & $^{+1.1}_{-0.1}$ \\
$\Delta q_3$     & $ 1.09$ & $^{+0.03}_{-0.02}$ & $^{+0.05}_{-0.05}$ & $^{+0.06}_{-0.01}$ & $ 1.07$ & $^{+0.03}_{-0.02}$ & $^{+0.06}_{-0.06}$ & $^{+0.10}_{-0.01}$ \\
$\Delta q_8$     & $ 0.30$ & $^{+0.06}_{-0.05}$ & $^{+0.05}_{-0.05}$ & $^{+0.23}_{-0.01}$ & $ 0.42$ & $^{+0.05}_{-0.08}$ & $^{+0.06}_{-0.06}$ & $^{+0.03}_{-0.01}$ \\
$\Delta \Sigma$  & $ 0.20$ & $^{+0.05}_{-0.06}$ & $^{+0.04}_{-0.05}$ & $^{+0.01}_{-0.01}$ & $ 0.25$ & $^{+0.07}_{-0.07}$ & $^{+0.05}_{-0.05}$ & $^{+0.05}_{-0.02}$ \\
$\Delta q_0$  & $ 0.20$ & $^{+0.05}_{-0.06}$ & $^{+0.04}_{-0.05}$ & $^{+0.01}_{-0.01}$ & $ 0.21$ & $^{+0.05}_{-0.06}$ & $^{+0.06}_{-0.07}$ & $^{+0.05}_{-0.02}$ \\
$\Gamma_1^p$     & $ 0.112$ & $^{+0.006}_{-0.006}$ & $^{+0.008}_{-0.008}$ & $^{+0.009}_{-0.001}$ & $ 0.114$ & $^{+0.005}_{-0.006}$ & $^{+0.010}_{-0.011}$ & $^{+0.001}_{-0.003}$ \\
$\Gamma_1^n$     & $-0.056$ & $^{+0.006}_{-0.007}$ & $^{+0.005}_{-0.006}$ & $^{+0.002}_{-0.001}$ & $-0.051$ & $^{+0.005}_{-0.006}$ & $^{+0.006}_{-0.007}$ & $^{+0.001}_{-0.012}$ \\
$\Gamma_1^d$     & $ 0.026$ & $^{+0.005}_{-0.006}$ & $^{+0.005}_{-0.006}$ & $^{+0.005}_{-0.001}$ & $ 0.029$ & $^{+0.004}_{-0.005}$ & $^{+0.007}_{-0.008}$ & $^{+0.001}_{-0.007}$ \\
$\Gamma_1^{p-n}$ & $ 0.168$ & $^{+0.005}_{-0.004}$ & $^{+0.008}_{-0.007}$ & $^{+0.007}_{-0.001}$ & $ 0.165$ & $^{+0.004}_{-0.004}$ & $^{+0.009}_{-0.009}$ & $^{+0.013}_{-0.001}$ \\
\hline\hline
\end{tabular}
\end{center}
\end{table}
\begin{table}
\caption{E154 results on $g_1^n$ at the $Q^2$ of the measurement for
each spectrometer.  Also shown are results for $g_1^n$ evolved to
$\langle Q^2\rangle = 5$ GeV$^2$ according to
Eq.~(\protect{\ref{eq:g1evolved}}). Errors were propagated as
described in the text.}
\label{tab:g1spec}
\begin{center}
\begin{tabular}{cccc}
\hline\hline
$x_i$ &  $Q^2_i$ 
&$g_1^n(x_i,Q^2_i)$ 
& $g_1^n(x_i,5~{\rm GeV}^2)$ \\
 &  GeV$^2$  & $\pm\ {\rm stat.}\ \pm\ {\rm syst.}$
& $\pm\ {\rm stat.}\ \pm\ {\rm syst.}\ \pm\ {\rm evol.}$ \\
\hline
\multicolumn{4}{c}{$2.75^{\circ}$ spectrometer} \\
\hline
$0.017 $&$\z1.2$&$ -0.351\pm0.115\pm0.109$&$ -0.421\pm0.115\pm0.113\pm0.016$\\
$0.024 $&$\z1.6$&$ -0.374\pm0.071\pm0.064$&$ -0.409\pm0.071\pm0.066\pm0.007$\\
$0.035 $&$\z2.0$&$ -0.289\pm0.061\pm0.038$&$ -0.304\pm0.061\pm0.039\pm0.005$\\
$0.049 $&$\z2.6$&$ -0.212\pm0.041\pm0.022$&$ -0.215\pm0.041\pm0.023\pm0.004$\\
$0.078 $&$\z3.3$&$ -0.119\pm0.031\pm0.013$&$ -0.117\pm0.031\pm0.013\pm0.002$\\
$0.123 $&$\z4.1$&$ -0.075\pm0.030\pm0.010$&$ -0.073\pm0.030\pm0.010\pm0.001$\\
$0.173 $&$\z4.6$&$ -0.070\pm0.033\pm0.010$&$ -0.069\pm0.033\pm0.010\pm0.001$\\
$0.241 $&$\z5.1$&$ -0.053\pm0.028\pm0.008$&$ -0.053\pm0.028\pm0.008\pm0.000$\\
$0.340 $&$\z5.5$&$\m0.002\pm0.036\pm0.004$&$\m0.001\pm0.036\pm0.004\pm0.000$\\
$0.423 $&$\z5.8$&$\m0.027\pm0.059\pm0.007$&$\m0.027\pm0.059\pm0.007\pm0.000$\\
\hline
\multicolumn{4}{c}{$5.5^{\circ}$ Spectrometer} \\ 
\hline
$0.084 $&$\z5.5$&$ -0.152\pm0.029\pm0.019$&$ -0.153\pm0.029\pm0.019\pm0.001$\\
$0.123 $&$\z7.2$&$ -0.117\pm0.017\pm0.013$&$ -0.121\pm0.017\pm0.013\pm0.002$\\
$0.172 $&$\z8.9$&$ -0.059\pm0.016\pm0.009$&$ -0.066\pm0.016\pm0.009\pm0.003$\\
$0.242 $&$ 10.7$&$ -0.040\pm0.012\pm0.006$&$ -0.047\pm0.012\pm0.006\pm0.003$\\
$0.342 $&$ 12.6$&$ -0.019\pm0.012\pm0.005$&$ -0.024\pm0.012\pm0.005\pm0.001$\\
$0.442 $&$ 13.8$&$ -0.009\pm0.012\pm0.003$&$ -0.011\pm0.012\pm0.003\pm0.001$\\
$0.564 $&$ 15.0$&$\m0.003\pm0.008\pm0.001$&$\m0.003\pm0.008\pm0.001\pm0.000$\\
\hline\hline
\end{tabular}
\end{center}
\end{table}
\begin{table}
\caption{Combined results on $g_1^n$ at the $Q^2$ of the measurement.
Also shown are results for $g_1^n$ evolved to
$\langle Q^2\rangle = 5$ GeV$^2$ according to
Eq.~(\protect{\ref{eq:g1evolved}}).}
\label{tab:g1comb}
\begin{center}
\begin{tabular}{cccc}
\hline\hline
$x_i$ &  $Q^2_i$ 
&$g_1^n(x_i,Q^2_i)$ 
& $g_1^n(x_i,5~{\rm GeV}^2)$ \\
 &  GeV$^2$  & $\pm\ {\rm stat.}\ \pm\ {\rm syst.}$  
& $\pm\ {\rm stat.}\ \pm\ {\rm syst.}\ \pm{\rm evol.}$ \\
\hline
$0.017$&$\z1.2$&$ -0.351\pm0.115\pm0.109$&$ -0.421\pm0.115\pm0.113\pm0.016$\\
$0.024$&$\z1.6$&$ -0.374\pm0.071\pm0.064$&$ -0.409\pm0.071\pm0.066\pm0.007$\\
$0.035$&$\z2.0$&$ -0.289\pm0.061\pm0.038$&$ -0.304\pm0.061\pm0.039\pm0.005$\\
$0.049$&$\z2.6$&$ -0.204\pm0.040\pm0.022$&$ -0.207\pm0.040\pm0.023\pm0.004$\\
$0.081$&$\z4.5$&$ -0.137\pm0.021\pm0.016$&$ -0.136\pm0.021\pm0.016\pm0.002$\\
$0.123$&$\z6.6$&$ -0.108\pm0.015\pm0.012$&$ -0.111\pm0.015\pm0.012\pm0.002$\\
$0.173$&$\z8.2$&$ -0.061\pm0.014\pm0.009$&$ -0.067\pm0.014\pm0.009\pm0.003$\\
$0.242$&$\z9.8$&$ -0.042\pm0.011\pm0.007$&$ -0.048\pm0.011\pm0.007\pm0.003$\\
$0.342$&$ 11.7$&$ -0.017\pm0.011\pm0.005$&$ -0.021\pm0.011\pm0.005\pm0.001$\\
$0.441$&$ 13.3$&$ -0.007\pm0.011\pm0.002$&$ -0.009\pm0.011\pm0.002\pm0.001$\\
$0.564$&$ 15.0$&$\m0.003\pm0.008\pm0.001$&$\m0.003\pm0.008\pm0.001\pm0.000$\\
\hline\hline
\end{tabular}
\end{center}
\end{table}
%

%
%

%

\end{document}